# TRANSPARENT CACHING OF VIRTUAL STUBS FOR IMPROVED PERFORMANCE IN UBIQUITOUS ENVIRONMENTS

Lachhman Das Dhomeja[1], Yasir Arfat Malkani[1], Asad Ali Shaikh[2] and Ayaz Keerio[3]

[1]School of Informatics, University of Sussex, Brighton, UK
`l.d.dhomeja@sussex.ac.uk` and `y.a.malkani@sussex.ac.uk`

[2]Institute of Information & Communication Technology, University of Sindh, Jamshoro, Pakistan
`asad.shaikh@usindh.edu.pk`

[3]Institute of Mathematics & Computer Science, University of Sindh, Jamshoro, Pakistan
`ayaz@usindh.edu.pk`

## ABSTRACT

*Context-awareness is an essential requirement for pervasive computing applications, which enables them to adapt and perform tasks based on context. One of the adaptive features of context-awareness is contextual reconfiguration. Contextual reconfiguration involves discovering remote service(s) based on context and binding them to the application components to realize new behaviors, which may be needed to satisfy user needs or to enrich user experience. One of the steps in the reconfiguration process involves a remote lookup to discover the service(s) based on context. This remote lookup process provides the largest contribution to reconfiguration time and this is due to fact that the remote calls are much slower than local calls. Consequently, it affects system performance. In pervasive computing applications, this may turn out to be undesirable in terms of user experience. Moreover, other distributed applications using the network may be affected as every remote method call decreases the amount of bandwidth available on the network. Various systems provide reconfiguration support and offer high-level reconfiguration directives to develop adaptive context-aware applications, but do not address this performance bottleneck. We address this issue and implement seamless caching of virtual stubs within our PCRA[1] for improved performance. In this paper we present and describe our transparent caching support and also provide its performance evaluation.*

---

[1] PCRA [1] is a policy-based context-aware adaptation system, which enables the development and execution of adaptive context-aware applications using Ponder2 [3,4] policy specifications.

                                        1

# 1. INTRODUCTION

Pervasive computing offers an environment where mobile and stationary computing devices are connected wirelessly or through wire, and interact with each other seamlessly to perform tasks in background to support us in our everyday life without requiring attention from us. In order for pervasive computing systems to be able to fulfill this vision, they need to adapt themselves in response to constantly changing conditions of a pervasive computing environment such as user context (user presence, user location, user activity, user preferences), varying network quality, mobility of users, device characteristics (CPU power, memory, battery power and small display). This makes context-awareness an essential requirement for pervasive computing systems.

There has been extensive research work in the field of context-awareness and a summary of contributions made in this field can be found in [5, 6]. Many researchers [5, 7, 8, 9] have attempted to identify and define core features of context-awareness. One of the core features identified by these researchers is a contextual reconfiguration. Contextual reconfiguration is a process of modifying or reconfiguring the structure of the application in response to contextual information to realize new behavior, which may be required to fulfill user needs or to enrich user experience. This can be realized by locating the remote services based on context and binding them to application component. For example, a user may want to have her messages displayed or printed to the nearest rendering device to her location. This requires discovering a device based on the location of the user and binding to it, and then sending messages to the bound device. As another example, contextual reconfiguration may be used to enrich the experience of a mobile user by providing her with a service of interest with respect to her changed location without requiring any cooperation from her. For instance, when the user is standing near a cinema, a movie information service could send information about the movies being exhibited in that cinema.

Our primary research objectives have been (1) providing a broader scope of adaptation by integrating various adaptation approaches to enable development of diverse set of adaptive context-aware applications and (2) providing a high-level programming model to simplify the development of adaptive context-aware applications. As a result, we have designed and implemented PCRA [1], which enables the development of adaptive context-aware applications using Ponder2 policy specifications. The broader scope of adaptation within PCRA is supported by integration of various adaptation types: Contextual Reconfiguration, Parameter Adaptation and Reconfiguration to managing invalid bindings.

One of the adaptation types supported by PCRA is Contextual Reconfiguration. Our approach to context-aware reconfiguration involves discovering remote service(s) based on contextual information and binding them to user instance, which represents the current user of the environment. Refer to [1] for the detailed description of our proposed approach to contextual reconfiguration. If we look at this binding process, one of the steps involved is performing a remote lookup to discover the remote service in response to context. This remote lookup process provides the largest contribution to the overall binding time. This is because the remote method calls are much slower than local calls, at least 1000 times slower [17]. In pervasive computing applications, it turns out to be undesirable in terms of user experience. For example, in the simple light example scenario, the user experience may get affected when the user enters room1 and light or air-conditioning in room1 is not turned on and adjusted to user's preferred value immediately. In the context of other distributed systems, this may affect other applications running on the network as every remote method call decreases the amount of bandwidth available on the network for all the applications using the network. To this end, we propose and implement a seamless caching technique to improve performance of pervasive computing applications.

There exists a large body of research work, such as [10, 11, 12, 13] which provide reconfiguration support and enable the development of adaptive context-aware applications using high-level reconfiguration directives, but do not address this performance bottleneck. We address this issue, and propose and implement caching technique within PCRA. In this paper we focus on PCRA's transparent caching support, while other features of PCRA, which include contextual reconfiguration, contextual adaptation and reconfiguration to manage invalid bindings, are out of the scope of this paper.

The remainder of this paper is structured as follows: section 2 presents a high-level design of PCRA and briefly describes its constituents. Section 3 presents and describes our caching support of virtual stubs integrated within PCRA. Section 4 provides performance evaluation of our caching support. Section 5 discusses some related work. In section 5, we conclude the paper.

## 2. HIGH-LEVEL SYSTEM ARCHITECTURE OF PCRA

PCRA is a policy-based context-aware adaptive system that allows developing and executing adaptive context-aware applications using Ponder2 policy specifications. In this section we present the high-level system architecture of PCRA and briefly provide an overview of all its parts. The high-level system design of PCRA is shown in the figure 1. PCRA architecture comprises three parts: Ponder2 system, Reconfiguration & Adaptation Infrastructure (RAI) and Remote Method Invocation (RMI). We describe each of its parts briefly.

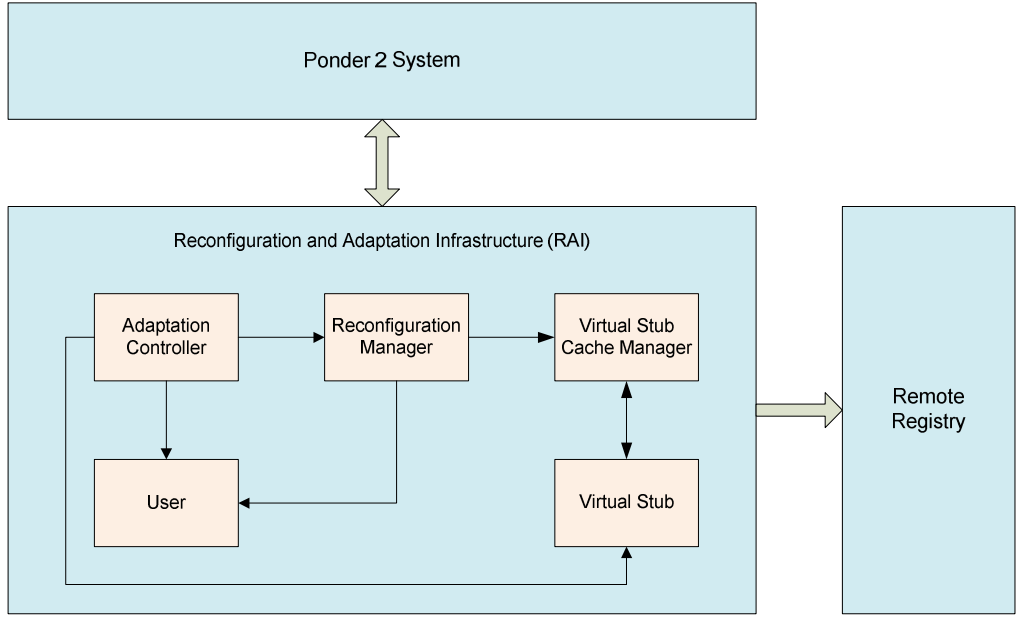

Figure 1: High-level System Architecture of PCRA

## 2.1 Ponder2

Ponder2 [3, 4] is a general purpose object management system, which is re-design and re-implementation of Ponder [14]. There exists a number of policy systems, such as [3, 14, 15, 18], but the reason for choosing and building our adaptation support on top of Ponder2 is that it is a light-weight, self-contained, extensible and scalable policy system for specifying and enforcing policies. These features make it suitable to be used on resource-constrained devices (e.g., PDA and mobile phones). Ponder2 provides the support for obligation policies and authorization policies and also includes other services such as domain service. The domain service allows grouping managed objects into hierarchical structure for management purposes. Ponder2 has an object language called PonderTalk [16] for configuring and controlling Ponder2 system, and Ponder2 obligation polices and authorization policies are expressed with PonderTalk commands. Adaptive context-aware applications are developed within PCRA by specifying behavior of applications using Ponder obligation policies.

## 2.2 Reconfiguration and Adaptation Infrastructure (RAI)

As discussed before, one of our research objectives has been the provision of wider scope of adaptation by integrating and providing runtime support for different adaptation types: Contextual Reconfiguration, Parametric Adaptation and reconfiguration to manage invalid bindings. We have designed and implemented RAI, which provides runtime support of these adaptation approaches. As mentioned before, the reconfiguration support involves remote lookups and this introduces performance bottleneck, which may affect user experience in pervasive computing environments. To address this issue, we have implemented caching support of virtual stubs within RAI. The design components of RAI, which provide the implementation of all features supported by RAI, are: Reconfiguration Manager, User Component, Virtual Stub, Virtual Stub Cache Manager and Adaptation Controller. The detailed description of these components can be found in [2]. In this paper, we only present and describe PCRA's caching support, while other features of PCRA are out of scope of this paper.

## 2.3 Java RMI

PCRA assumes RMI services to be available in the environment, so the developers of remote services in PCRA environment use RMI infrastructure to develop remote services and register them in the RMI registry. As a part of reconfiguration process, PCRA uses RMI infrastructure to discover these services based on context and bind them to the current user of the environment.

To summarize, these three parts of PCRA architecture (Ponder2, RAI and RMIA) work together to provide a policy-based development platform, in which adaptive context-aware applications are developed using policies. Ponder2 provides infrastructure for specification and enforcement of policies while RAI provides runtime adaptation support. In addition to its runtime adaptation support, RAI also includes caching support for improved performance. In next section, we present design and description of caching support integrated within RAI.

## 3. SEAMLESS CACHING SUPPORT FOR IMPROVED PERFORMANCE

As virtual stub is the one which is cached, we first briefly describe the virtual stub. The virtual stub wraps the real proxy of remote service and any call to this remote service is intercepted by the virtual stub and then forwarded to the remote service. In addition to forwarding the remote call, it handles the issue of invalidity of bindings. Refer to [2] for the detailed description of virtual stub and our transparent reconfiguration approach to managing invalid bindings.

In our approach to contextual reconfiguration, the remote service is discovered based on context, an instance of virtual stub is created and initialized with the real proxy of found service, cached locally and then the virtual stub is handed to the user instance. Once the user instance has the virtual stub, this means that user has the binding to the remote service through its corresponding virtual stub. User component is one of the design components of RAI as described in section 2, which models the user of the environment and holds the bindings to remote services. When the binding to the same services are needed again, the virtual stubs for these services are obtained from local cache, thus avoiding remote lookups for improved performance. During the binding process when the system discovers a particular remote service for first time, it takes longer since it involves a remote lookup. When the binding to the same service is required again, the system takes less time since the service (the virtual stub holding this service) is obtained from the local cache directly without the need for a remote call. Consider a simple example scenario in which bindings to light service and air-conditioning service in room1 are created upon the entrance of the user in room1 and the light service and the air-conditioning services are turned on, and when the user leaves room1, these are turned off. When the first user enters room1, the binding time taken by the system would be far higher as the light service and air-conditioning services are discovered by performing remote lookups. When this user leaves room1, the light service and air-conditioning services are turned off and after some time other user enters the room1, the system would create the bindings between the user instance for the new user and both the light service and air-conditioning in the room1, and the binding time taken by the system would be greatly reduced in comparison to what it would be in the former case since both the virtual stub for light service and the virtual stub for air-conditioning services are obtained locally. In performance evaluation section, we show that the caching technique significantly reduces reconfiguration time and improves system scalability.

The design component in charge of providing support for seamless caching of virtual stubs is the virtual stub cache manager. As a part of reconfiguration process, the reconfiguration manager interacts with the virtual stub cache manager to support in the process of establishing bindings to remote services. In the process, the virtual stub cache manager is involved in caching virtual stubs, in addition to performing other operations. Figure 2 shows the interaction of virtual stub cache manager with other design components of RAI to realize caching support of virtual stubs, and the figure 3 shows the partial implementation of virtual stub cache manager.

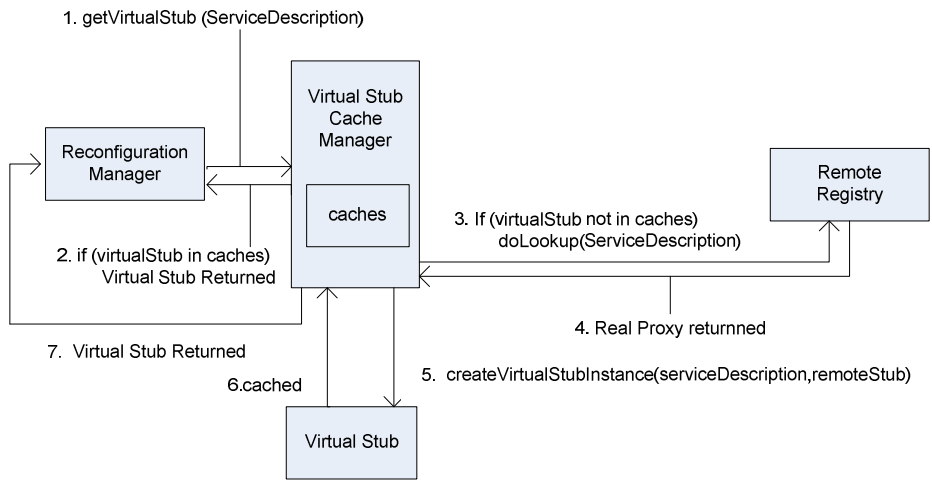

Figure 2: Seamless caching of virtual stubs

The description of steps involved in figure 2 above is given below.

- The reconfiguration manager communicates with the virtual stub cache manger by invoking getVirtualStub(serviceDescription) of virtual stub cache manager (figure 3, line 3) and this method returns the virtual stub.
- The virtual stub cache manager has a hashtable named virtualStub_Table (figure 3, line 2) where virtual stubs for different services are cached. In order to return the requested virtual stub to the reconfiguration manager, it checks whether it is in virtualStub_Table by invoking virtualStub_Table.get(serviceDescription) (figure 3, line 5). If available, it is returned to the reconfiguration manager (figure 3, lines 6, 7).
- If the virtual stub cache manager does not find it, it invokes its doLookup(serviceDescription) method (figure 3, line 9) and this, in turn, communicates with the remote registry to obtain the real proxy/stub of the remote service (figure 3, line 19).
- After obtaining the real proxy of the service, the virtual stub cache manager creates an instance of the virtual stub and initializes it with the found proxy through createVirtualStubInstance (serviceDescription, remoteStub) method (figure 3, line 20).
- The instance of virtual stub is then cached into the table through a method call virtualStub_Table.put (serviceDescription, virtualStub) (figure 3, line 22) and then returned to the reconfiguration manager(figure 3**,** line 23) .

International Journal of UbiComp (IJU), Vol.2, No.4, October 2011

```
1.  public class VirtualStubCache {
2.  static Hashtable<String, String> virtualStub_Table = new Hashtable<String, String>();
3.  String getVirtualStub (String servName) {
4.  String virtualStubString=null;
    ….
5.  virtualStubString = (String)virtualStub_Table.get(serviceName);
6.  if(virtualStubString!=null){
7.  return virtualStubString;
8.  } else{
9.  virtualStubString=doLookup(servName);
10. return virtualStubString;
11. }
12. virtualStubString=doLookup(servName);
13. return virtualStubString;
14. }

15. String doLookup(String servName){
16. VirtualStub vs=null;
17. String serializedVirtualStub;
18. try {
19. Remote rs=  Naming.lookup("rmi://192.168.1.1/" + servName);
        ………
        ………
20. vs = createVirtualStubInstance(servName,rs);
21. serializedVirtualStub= serializeVirtualStubIntoString(vs);
22. virtualStub_Table.put(serviceName, serializedVirtualStub);
23. return serializedVirtualStub;
24. } catch (MalformedURLException e) {
25. }
    …..
    …..
26. }

27. VirtualStub createVirtualStubInstance(String serviceName, Remote rs){
28. return new VirtualStub(serviceName,rs);
29.  }
        ……
        ……

30. }
```

Figure 3: Partial source code of virtual stub cache manager

The cached virtual stub may become outdated when the real proxy held by the cached virtual stub becomes invalid. This requires updating the cached virtual stub. The real proxy of the remote service may become invalid due to various reasons (e.g., due to power failures on hosting device where the remote service is running or the service has been moved to some other location, etc). Any method call on an invalid proxy would result in an exception being thrown and caught by the virtual stub. As a response to this exception, the virtual stub discovers a new copy of real proxy and replaces the old copy with the new one. Now the virtual stub has the updated copy of real proxy, but the cached copy of virtual stub contains the invalid proxy. The updated virtual stub sends a copy of itself to the virtual stub cache manager and then the virtual stub cache manager replaces the old copy with new one in the cache.

## 4. PERFORMANCE EVALUATION

In order to study the performance of our caching support, we measure reconfiguration time without using cache and reconfiguration time using cache. The reconfiguration time is the time taken by PCRA when creating a binding between a user instance and a remote service in response to policy evaluation. The reconfiguration time is measured from a point context is sent by the context monitor to the policy, which has subscribed to it, until the binding is established in response to policy evaluation. Figure 4 shows sequence diagram for reconfiguration time without using cache.

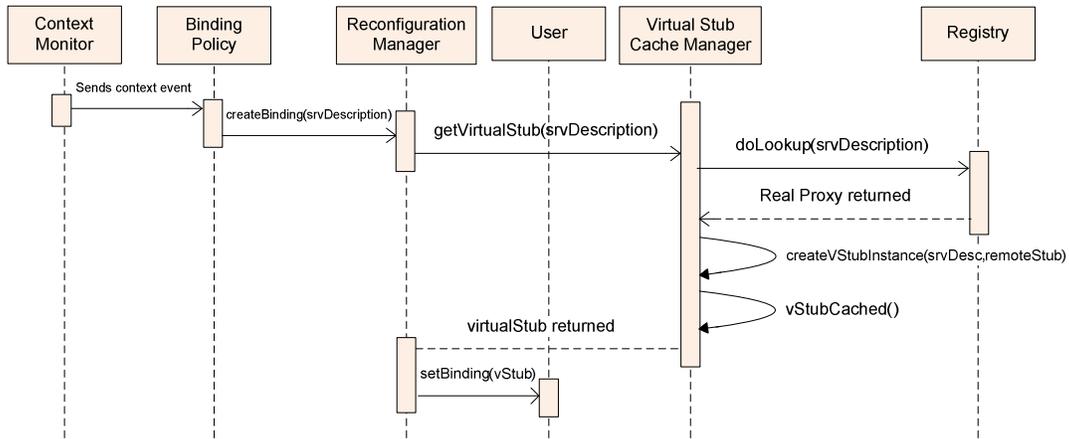

Figure 4: Sequence diagram for reconfiguration time without cache

As can be noted from figure 4, the remote service is discovered through a remote call to the registry and a virtual stub instance for this service is created and initialized with the corresponding real proxy and given to a user instance. Whereas reconfiguration time using cache involves obtaining the virtual stub for the required service from the local cache directly without the need for a remote call and giving it to the user instance, which is shown in the figure 5.

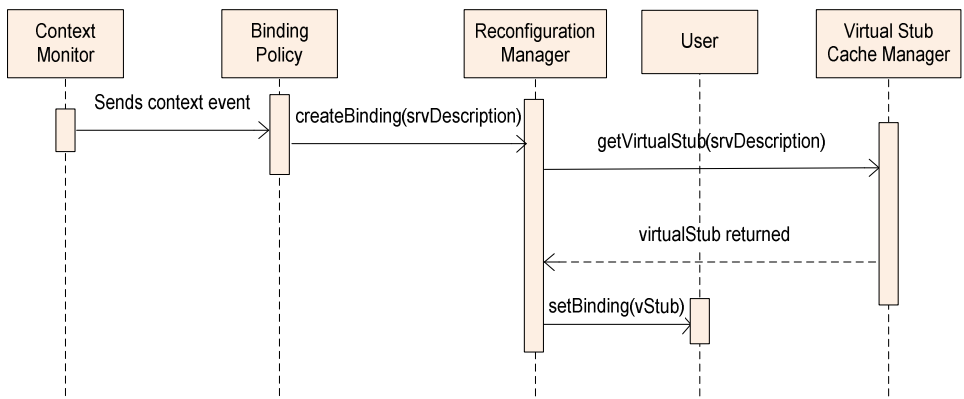

Figure 5: Reconfiguration time with cache sequence diagram

## 4.1 Experimental Setup

We conducted tests to measure reconfiguration time without using cache and using cache. We used four machines and ran tests on each of them. The four machines used were: (1) Intel Celeron 501 MHz, 256MB RAM; (2) Intel Pentium 3 734 MHz CPU, 384MB RAM; (3) Intel P4 2.4GHz, 1GB RAM; and (4) Intel Xeon, 2 GHz, 2GB RAM. All machines were running windows XP and used JDK1.5. The JDK1.5 or above is the requirement for running PCRA. As can be noted, machines used range from a less powerful machine to a typical desktop PC. The purpose of running tests on each of these machines was to investigate how PCRA scales from less powerful machines to powerful desktops. In configuration setup, PCRA (Ponder2 system and Reconfiguration and Adaptation Infrastructure (RAI)), RMI registry, remote services were running on the same machine.

## 4.2 Test Results

**Test 1:** Reconfiguration time without cache

To measure this time on each of four machines, the binding policy was triggered 20 times. This policy responds to a user presence event and its action part has a reconfiguration message to discover and bind the light service to the user instance.

**Test 2:** Reconfiguration time with cache

To measure this time on each of four machines, the same binding policy used in Test1 was triggered 20 times. In the binding process, the virtual stub was obtained from the local cache directly without the need for a remote call and delivered to the user instance.

**Results:** The reported times for test 1 and test 2 are average times and are presented graphically in figure 6 along with standard deviation. Reconfiguration time includes a RMI lookup time (time to discover a remote service) and this provides the largest contribution to reconfiguration time. This is due to fact that the remote calls are much slower than local calls, at least 1000 times slower. As can be noted, reconfiguration time with caches on each of machines is a far lower than the reconfiguration time without caches. This clearly shows that the seamless caching support of virtual stubs provided by PCRA significantly reduces reconfiguration time (hence improves performance) and improves system scalability. Note that the binding policy triggered in test 1 involves discovering only one remote service, light service (i.e., the user has a binding with one service), thus reconfiguration time includes just one remote lookup time. However, the user can have multiple bindings and each binding involves a remote lookup time. As a result, the reconfiguration time with each additional binding would increase significantly. As expected, as shown in figure 6, reconfiguration time with and without caches from a less powerful machine to a powerful machine reduces considerably and vice versa. This indicates that PCRA can scale down and run on resource-constrained devices.

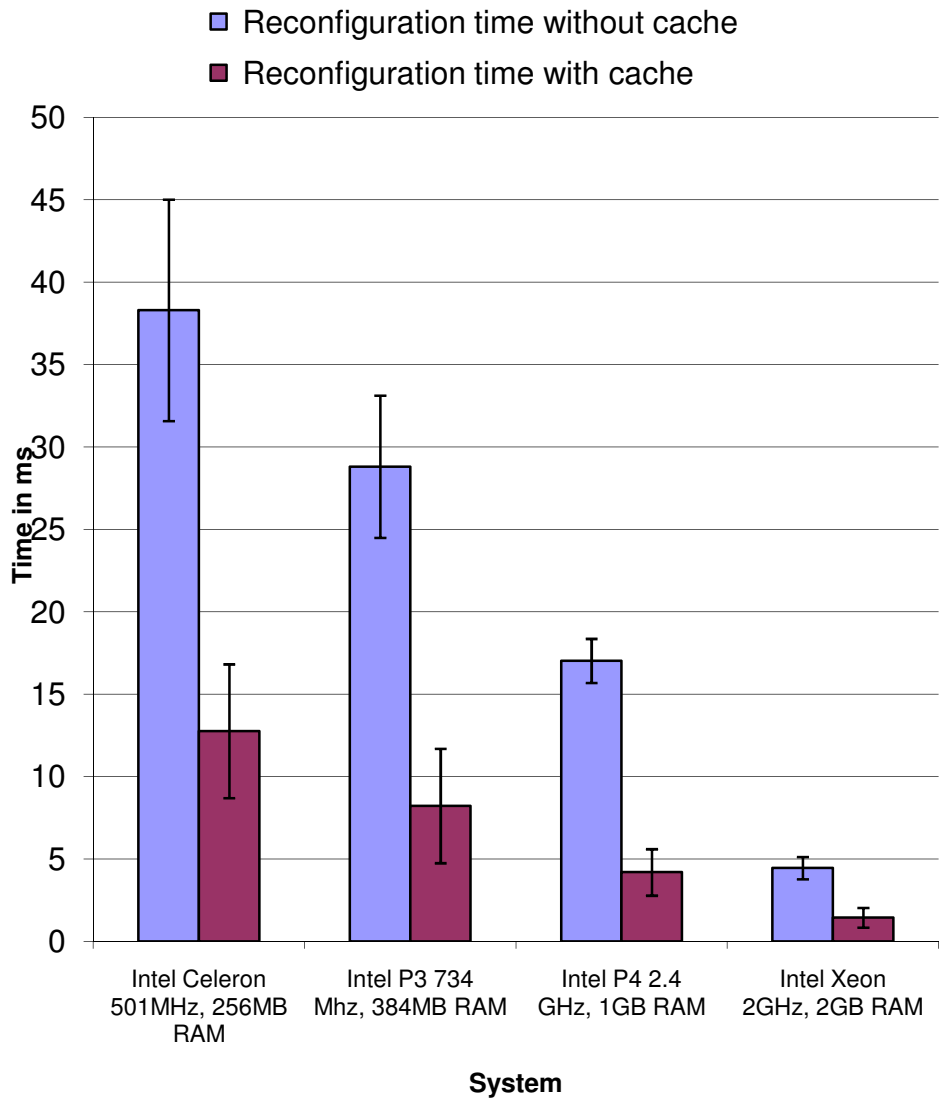

Figure 6: Reconfiguration time comparisons with and without cache

## 5. RELATED WORK

There exists a large body of work on caching, especially on web caching, in which various caching techniques have been deeply examined and applied [24, 25]. Other environments such as mobile and ubiquitous environments are dynamic, where dynamicity arises out of various factors such as weak connectivity (resulting from intermittent communication due to low bandwidth, high latency, etc), frequent disconnections and heterogeneity of mobile clients. Weak connectivity and frequent disconnection problems affect the user experience. To address these issues and hence to improve user experience, various caching techniques have been proposed and implemented in mobile and ubiquitous environments. In this section we focus on providing related work in mobile environments and compare it with the proposed caching support.

Various research efforts [19, 20, 21, 22, 23, 26] target mobile environments and support disconnections, which enable mobile clients to carry on accessing services in case of disconnections. The system [19] provides support for caching of web services for supporting disconnections in mobile environments, where SOAP (Simple Object Access Protocol) requests / responses are cached. This system involves a design component called SOAP proxy, which caches SOAP request sent by the client and the SOAP response made by the service. When the client is disconnected and requests to access the service, the SOAP proxy obtains the response from the cache (if available in the cache) and serves it to the mobile host. Other systems that support caching of webs services for mobility include [20, 23].

The middleware called SCaLaDE [21] targets mobile environments and provides the support for rebinding of links to information resources in response to mobile entity migration. In addition to this, SCaLaDE also supports disconnection operations during which service request are carried on. SCaLaDE uses a design component, called mobile proxy, which is responsible to perform all kinds of adaptation supported by SCaLaDE including handling of disconnections.

The authors in [22] have developed a framework called Domint, which supports disconnections in CORBA environment. Domint enables mobile hosts to carry on accessing distributed CORBA objects when the mobile hosts are disconnected. The working of Domint involves intercepting request made to CORBA ORB and then transparently rerouting these requests to the local disconnected object.

It can be noted that the above research efforts target mobile environments and handle temporary disconnections to improve mobile user experience by caching request / response and serving these cached responses to mobile clients when disconnected. In contrast to these systems, we target smart environments (e.g., smart homes) where our caching support of virtual stubs improves the experience of users of the smart environment. The virtual stub wraps a real proxy of the remote service and in PCRA the binding to the remote service is through its corresponding virtual stub. As a part of our caching support, the virtual stub is cached locally and when the binding to the same remote service is required again, the virtual stub is obtained from local cache without the need for a remote lookup, hence improving the user experience.

## 6. CONCLUSION

User experience is an important factor to consider while designing and implementing systems that enable development and execution of mobile and ubiquitous computing applications. In mobile environments, user experience is affected due to dynamicity of the mobile environments (e.g., weak connectivity, frequent disconnections, etc) while accessing services. To improve the mobile user experience, various caching techniques have been proposed, which include caching request / response pair (request made by the mobile client and the response made by the service) and local availability of the remote object. In other ubiquitous computing environments such as smart homes, realization of many applications require discovering remote services based on context and then binding them to application component. Discovery of remote service based on context requires remote lookups, which provide a largest contribution to overall binding time. This results in system performance degradation, thus affecting user experience. In this paper we presented and described our caching support of virtual stubs to address this issue. The key design component of caching support is the virtual stub cache manager, which interacts with other RAI design components to provide caching support. As a part of reconfiguration support provided by PCRA for establishing bindings with the remote services, the remote service was discovered based on context, an instance of virtual stub was created and initialized with the real proxy of the remote service and then it was locally cached. When the binding to the same service was needed again, its corresponding virtual stub was obtained from the local cache without the need for performing a lookup to discover the required service, thus improving application performance. We also provided performance evaluation of the proposed caching support, which clearly indicates that it significantly reduces reconfiguration time, hence improved user experience.

## Authors

**Dr. Lachhman Das Dhomeja** is an Assistant Professor at the Institute of Information & Communication Technology (IICT), University of Sindh, Jamshoro, Pakistan. He got his Master's degree in Computer Technology from University of Sindh, Jamshoro (Pakistan) in 1991 and PhD from University of Sussex, UK in 2011. His main research area is Pervasive Computing in general and policy-based context-awareness in particular. His other research interests include secure device pairing in ubiquitous environments, software architectures and Distributed Computing.

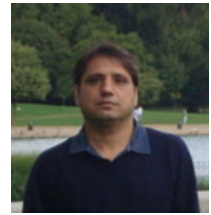

**Dr. Yasir Arfat Malkani** is a Lecturer at the Institute of Mathematics and Computer Science (IMCS), University of Sindh, Jamshoro, Pakistan. He go his Master's degree in Computer Science from University of Sindh, Jamshoro (Pakistan) in 2003 and PhD from University of Sussex, Brighton, UK in 2011. His main area of research is Pervasive Computing. His research is focused on secure device/service discovery and access control mechanisms using policies and location/proximity data/information. He is also interested in sensor networks, wireless networks (including WiFi, Bluetooth, WiMAX, etc), and solutions to various issues in distributed and pervasive computing systems through the integration of tools and techniques from distinct disciplines/areas.

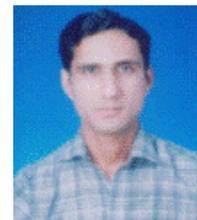

**Dr. Asad Ali Shaikh** is an Associate Professor and director of the Institute of Information and Communication Technology (IICT), University of Sindh, Jamshoro, Pakistan. He did his Masters degree in Computers Engineering from Clarkson University, USA in 1991 and PhD degree in Information Technology from University of Sindh, Pakistan in 2006. His current research focus is on the protocol design, security issues in computer networks and software development. 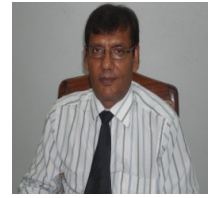

**Dr. Ayaz Keerio** is an Assistant Professor at the Institute of Mathematics and Computer Science (IMCS), University of Sindh, Jamshoro, Pakistan. He got his Master's degree in Computer Science from University of Sindh, Jamshoro (Pakistan) and PhD from University of Sussex, UK in 2011. His main area of research is Speech Synthesis and Recognition. He is also interested in computer networks and mobile & distributed computing systems. 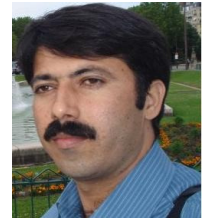